\documentclass[epj]{webofc}
\usepackage[varg]{txfonts}   
\usepackage{xcolor}
\wocname{ }
\woctitle{ }
\newcommand{\beq}{\begin{equation}}
\newcommand{\eeq}{\end{equation}}

\newcommand{\beqa}{\begin{eqnarray}}
\newcommand{\eeqa}{\end{eqnarray}}

\newcommand{\p}{\partial}

\begin{document}
\selectlanguage{english}
\title{Instantaneous Dynamics of QCD}
%
%

\author{Patrick Cooper\inst{1}\fnsep\thanks{\email{cooperp@duq.edu}} \and
        Daniel Zwanziger\inst{2}\fnsep\thanks{\email{dz2@nyu.edu}} \hspace{-0.1cm}$^,\;$\footnote{Speaker, XIIth Quark Confinement and the Hadron Spectrum, 28 August to 4 September, 2016, Thessaloniki, Greece}  
}

\institute{Physics Department, Duquesne University, 600 Forbes Ave, Pittsburgh, PA 15282, USA 
\and
           Physics Department, New York University, 4 Washington Place, New York, NY 10003, USA       
}

\abstract{%
  We start from the observation that, in the confining phase of QCD, the instantaneous color-Coulomb potential in Coulomb gauge is confining.  This suggests that, in the confining phase, the dynamics, as expressed in the set of Schwinger-Dyson equations, may be dominated by the purely instantaneous terms.  We develop a calculational scheme that expresses the instantaneous dynamics in the local formulation of QCD that includes a cut-off at the Gribov horizon.\footnote{This talk is dedicated to the memory of Martin Schaden.}
}
\maketitle


\section{Introduction}\label{intro}

In sections 1 through 9 we outline the properties and derivation of the instantaneous dynamics expressed in the instantaneous Schwinger-Dyson equations (ISDE).  In sect.\ 10, we write the ISDE in diagrammatic form.  

\section{No confinement without Coulomb confinement} 
 
 Gribov's insight into the mechanism of confinement \cite{Gribov:1977wm} is substantiated by the theorem \cite{Zwanziger:2002},
\beq
V_{\rm coul}(R) \geq V_{\rm wilson}(R)
\eeq
for $R \to \infty$, where the color-Coulomb potential, $V_{\rm coul}(R)$, is the instantaneous part of the zero-zero component of the gluon propagator in Coulomb gauge,\footnote{It is shown in \cite{Zwanziger:1998} that $ V_{\rm coul}(R)$ and $gA_0$ are renormalization-group invariants in Coulomb gauge.}
\beq
g^2 D_{00}(R, T) = V_{\rm coul}(R) \ \delta(T) + \mathrm{non-instantaneous},
\eeq
and $V_{\rm wilson}(R)$ is the gauge-invariant potential derived from a large rectangular Wilson loop.  Accordingly, when $V_{\rm wilson}(R)$ rises linearly, $V_{\rm coul}(R)$ rises linearly or super-linearly.

\section{Motive and method of present approach}

Motivated by this theorem, we conjecture that, in the Coulomb gauge, the instantaneous dynamics is dominant.  This will be expressed in a closed system of instantaneous Schwinger-Dyson equations (ISDE).

We shall work within the framework of local quantum field theory, with a local action that  encodes the cut-off at the Gribov horizon  \cite{Zwanziger:1989mf}, and moreover that is BRST-invariant, thus preserving the geometric property of a gauge theory.  This action is derived by the method of Maggiore-Schaden \cite{Schaden:1994} in which  BRST-symmetry is spontaneously broken \cite{Schaden:1996}.  A conjecture is offered in \cite{Schaden:1412, Schaden:1501} for the identification of physical operators and the physical subspace, but this important subject will not be discussed here.

We shall derive the ISDE by extending to the local action that enforces the cut-off at the Gribov horizon the method which was previously applied to the Faddeev-Popov action~\cite{Alkofer:2009dm}.

\section{Horizon function and non-local action}

We shall localize the non-local Euclidean action,
\beq
S = S_{\rm FP} + \gamma H - \gamma \int d^dx d(N^2 -1),
\eeq
where $S_{\rm FP}$ is the (local) Faddeev-Popov action in Coulomb gauge, $H$ is the non-local horizon function given by
\beq
H \equiv  \int d^dx d^dy \ D_\mu^{ab}(x) D_\mu^{ac}(y) (M^{-1})^{bc}(x, y; A),
\eeq
and $\gamma^{1/4}$ is the Gribov mass. $H$ cuts off the functional integral at the Gribov horizon, as one sees from the eigenfunction expansion,
\beq
(M^{-1})^{bc}(x, y; A) = \sum_n {\psi_n^b(x) \psi_n^c(y) \over \lambda_n(A)},
\eeq
where the $\lambda_n(A)$ are the eigenvalues of the Faddeev-Popov operator $M \equiv - {\bf \nabla} \cdot {\bf D(A)}$ in Coulomb gauge.  The lowest non-trivial eigenvalue $\lambda_1(A)$ approaches 0 as the Gribov horizon is approached from within.

\section{Horizon condition and Kugo-Ojima confinement condition}

The Gribov mass $\gamma^{1/4}$ is not an independent parameter, but is fixed by the horizon condition
\beq
< H > = d (N^2 -1) \int d^dx.
\eeq
It is a remarkable fact that the horizon condition and the famous Kugo-Ojima confinement condition \cite{Kugo-Ojima, Kugo:1995km} are the same statement
\beq
-i \int d^dx \ < (D_\mu c)^a(x) (D_\mu \bar c)^a(0) > = d(N^2 - 1).
\eeq
This may indicate that color confinement is assured in this theory, although the precise hypotheses of the Kugo-Ojima theorem are not satisfied in this approach.

\section{Horizon condition and dual Meissner effect}

It is also a remarkable fact that the horizon condition is equivalent to the statement that the QCD vacuum is a perfect color-electric superconductor, which is the dual Meissner effect \cite{Reinhardt:2008},
\beq
G(\vec k) = {d(\vec k) \over \vec k^2 } = {1 \over \epsilon(\vec k) \vec k^2 }
\eeq
\beq
d^{-1}(\vec k = 0) = 0 \Longleftrightarrow \epsilon(\vec k = 0),
\eeq
where $G(\vec k)$ is the ghost propagator and $\epsilon(\vec k)$ is the dielectric constant.

\section{Auxiliary ghosts}

Just as the Faddeev-Popov determinant is localized by introducing ghosts,
\beq
\det M = \int dc d \bar c \exp\left( - \int d^dx \ \bar c M c \right),
\eeq
likewise, the horizon function $H$ may be localized by introducing ``auxiliary" ghosts \cite{Zwanziger:1989mf},
\beq
\exp( - \gamma H) = \int d\varphi d \bar\varphi d \omega d \bar\omega \exp\left(- \int d^dx \  \left[ \bar\varphi M \varphi - \bar\omega M \omega + \gamma^{1/2} D \cdot (\varphi - \bar\varphi ) \right] \right).
\eeq
For reviews of this approach, see \cite{Vandersickel:2012, Sobreira:2004}.

\section{Cancellation of energy divergences}

The Coulomb gauge is plagued by energy divergences.  For example, the integrand of the ghost loop is independent of $p_0$,
\beq
\int d^3p dp_0 {1 \over \vec p^2} {1 \over (\vec p - \vec k)^2},
\eeq
which leads to the horrible energy divergence
\beq
\int dp_0 \ 1 = \infty.
\eeq

We get rid of such divergences by using the first-order formalism in which they cancel manifestly \cite{Zwanziger:1998, Andrasi:2014}.  This relies on the fact that the Coulomb gauge is a unitary gauge.  The first-order action is obtained by writing  
\beq
\exp \left( - \int d^4x \ F_{0i}^2/2 \right) = \int d^3 \pi \ \exp \left( \int d^4x \ ( i \pi_i F_{0i} - \pi^2/2) \right).
\eeq
Here $F_{0i} = \p_0 A_i - D_i A_0$, the Coulomb gauge condition $\p_i A_i = 0$ holds identically, and $\pi_i$ is color-electric field.  It is decomposed into its transverse and longitudinal parts, 
\beq
\pi_i = \tau_i - \p_i \lambda, 
\eeq
with $\p_i \tau_i = 0$.  As a result of the equality  \eqref{equalprops}, given below, the $c - \bar c$ ghost loop cancels the $\lambda-A_0$ loop,
\beq
\int d^{d+1}p \ \left[ D_{A_0 \lambda}(\vec p) D_{A_0 \lambda}(\vec k + \vec p) - D_{c \bar c}(\vec p) D_{c \bar c}(\vec k + \vec p)
 \right] = 0,
 \eeq
and likewise for the pairs of auxiliary ghosts.  This cancellation eliminates the unwanted energy divergences.

\section{Local action and physical degrees of freedom in Coulomb gauge}

The local action is given by\footnote{For the action at finite temperature, see \cite{Cooper:1512.08}.}
\beq
S = \int d^{d+1}x \ (\mathcal L_1 + \mathcal L_2 + \mathcal L_3)
\eeq
\beq
\mathcal L_1 = i \tau_i D_0 A_i + (1/2)[ \tau^2 + (\p \lambda)^2 + B^2]
\eeq
\beq
\mathcal L_2 = i \p_i \lambda D_i A_0 - \p_i \bar c D_i c + \p_i \overline \varphi_j D_i \varphi_j - \p_i \overline \varphi_j D_i \varphi_j - \p_i \overline \omega_j D_i \omega_j
\eeq
\beq
\mathcal L_3 =  \gamma^{1/2} g f^{abc} A_j^b (\varphi - \overline \varphi)_{j a}^c,
\eeq
and $B_i$ is the color-magnetic field.  Only the first term, $i \tau_i \p_0 A_i$, contains a time derivative.  Because $A_i$ is identically transverse, $\p_i A_i = 0$ (as is $\tau_i$), $\p_0$ acts on the two would-be physical degrees of freedom.  The remaining terms impose constraints in the local theory.  The  energy divergences cancel between pairs of fermi and bose ghost loops, including the first pair, $i \p_i \lambda D_i A_0 - \p_i \bar c D_i c$.  The last term, $\mathcal L_3$, mixes bose-ghost and gluon fields.  

In the following it will be convenient to change variables from $\varphi$ and $\bar\varphi$ to $U$ and $V$ defined by
\beq
\varphi = (U + i V)/\sqrt 2; \hspace{2cm} \overline\varphi  = (U - i V)/\sqrt 2.
\eeq
The $V$-field mixes with the gluon field $A$, whereas the $U$-field does not.

The propagators of the fields $\lambda$ and $A_0$ are related to the ghost propagator $D_{c \bar c}$ by
\begin{equation}
\left(
\begin{array}{rlc}
D_{\lambda \lambda} & D_{\lambda A_0} \\
D_{A_0 \lambda} & D_{A_0 A_0}
\end{array}
\right)^{ab}
= \delta^{ab}
\left(
\begin{array}{rlc}
0 \ \ \ \  & \ \ - i D_{c \bar c}(\vec k)  \\
- i D_{c \bar c}(\vec k)   & \ \  \Gamma_{\lambda \lambda}(\vec k) D_{c \bar c}^2(\vec k)
\end{array}
\right),
\end{equation}
where $\Gamma_{\lambda \lambda}$ is the 2-point vertex function.  The equality of the Faddeev-Popov ghost propagator and the bose-ghost propagator,  
\beq
\label{equalprops}
\Gamma_{A_0 \lambda}^{-1} = D_{\lambda A_0} = - i D_{c \bar c}(\vec k).
\eeq
assures that the corresponding fermi and bose loops yield Faddeev-Popov determinants that cancel exactly,
\beq
{D(M) \over D(M)} = 1.
\eeq

\section{Instantaneous Schwinger-Dyson equations (ISDE)}

\begin{figure}
        \centering
        \includegraphics[width=10cm]{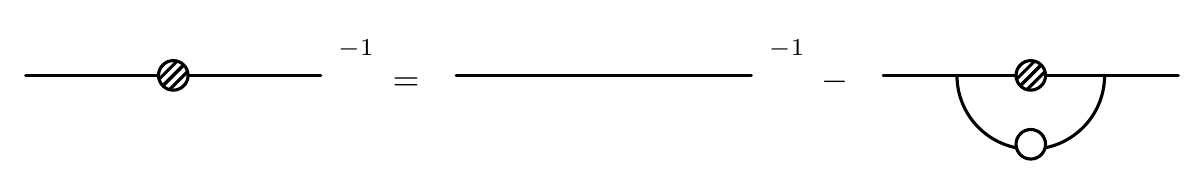}
        \caption{Schematic form of the ISDE. The shaded circle
        represents an instantaneous dressed propagator and the empty circle a non-instantaneous dressed propagator.  All three-point vertices are tree-level.}
        \label{generic}
\end{figure}

The truncation scheme which we use is represented schematically in Fig.~1, with detailed diagrams given in Figs.\ 2 and 3.  Propagators such as $D_{\lambda A_0}$ and $D_{AV}$ represent mixing.

\begin{figure}
        \centering
        \includegraphics[width=15cm]{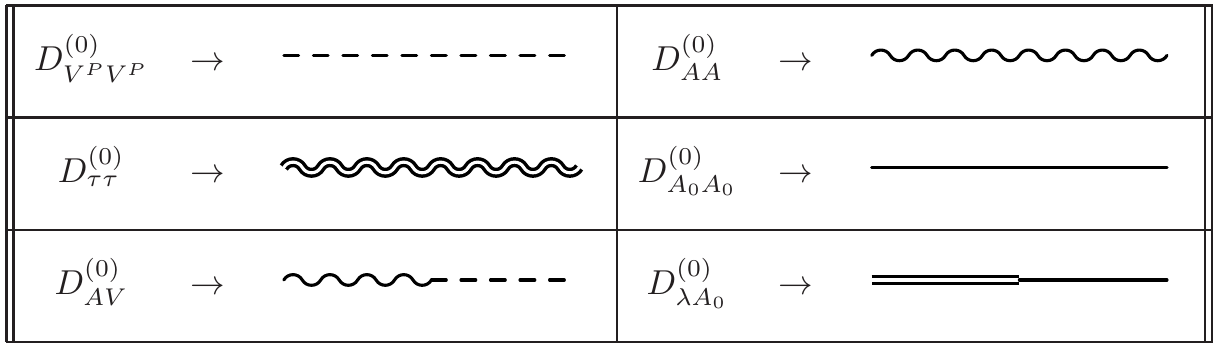}
        \caption{Free propagators that appear in the ISDE.  $V^P$ is the part of the bose-ghost propagator that mixes with the gluon propagator.}
\end{figure}

\begin{figure}
        \centering
        \includegraphics[width=15cm]{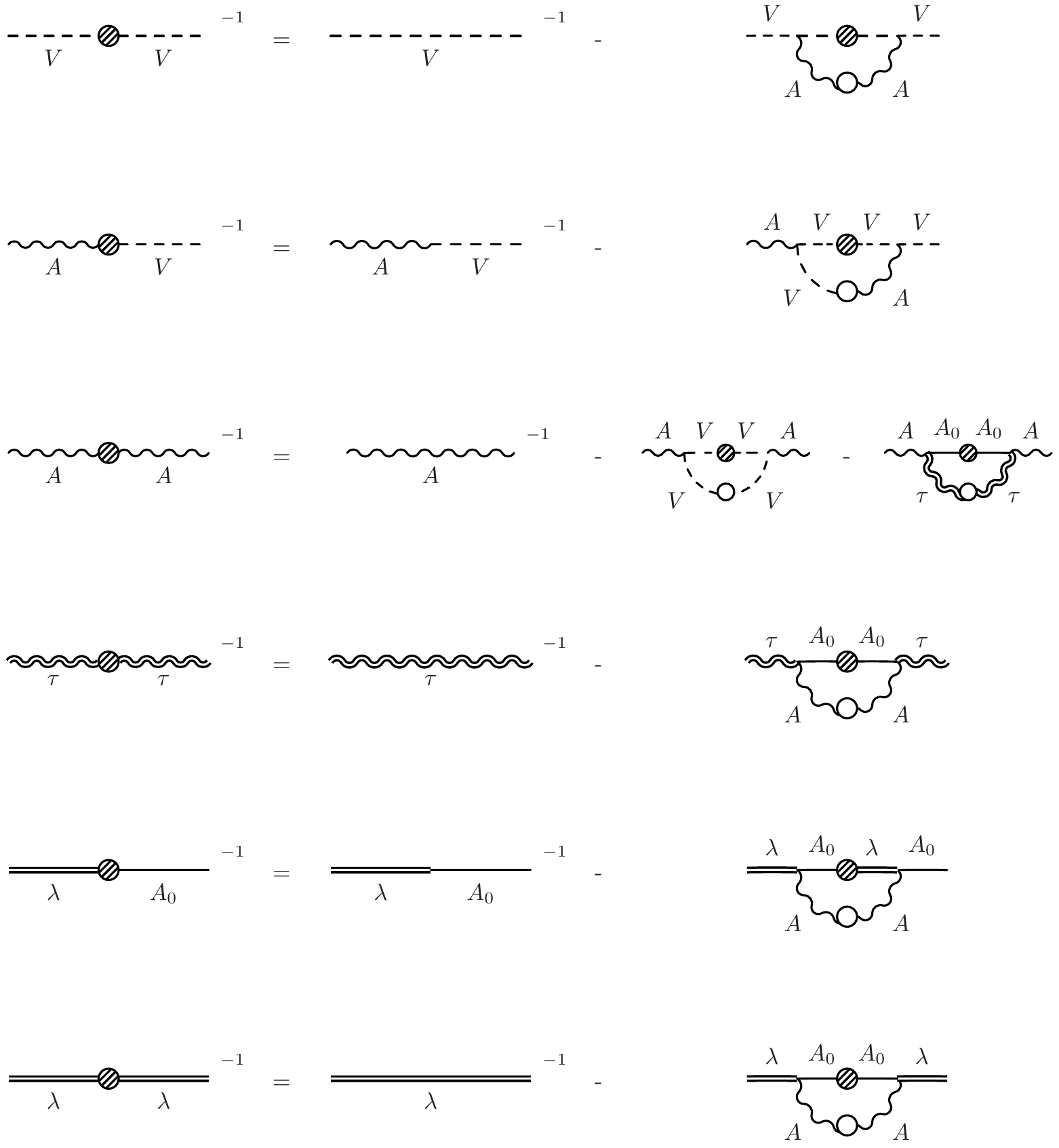}
        \caption{Diagrams for the relevant ISDEs. The five
        different types of lines are explained in Fig.\ 2. As usual,
        shaded circles represent the instantaneous, full propagator and the empty
        circles are non-instantaneous.}
        \label{equations}
\end{figure}

In Coulomb gauge, the propagators in general decompose into an instantaneous part and a non-instantaneous part,
\beq
D(\vec x, x_0) = D_I(\vec x) \delta(x_0) + D_N(\vec x, x_0)
\eeq
\beq
D(\vec p, p_0) = D_I(\vec p) + D_N(\vec p, p_0),
\eeq
where $\lim_{p_0 \to \infty} D_N(\vec p, p_0) = 0$.  The ISDE is obtained as follows.\\
\\
I.  Cancel loops with two instantaneous bose-ghost propagators against similar fermi-ghost loops, to get rid of energy divergences,\footnote{The part of the bose-ghost loop, that is caused by mixing with the gluon propagator, survives.}
\beq
\int d^{d+1}p \ \left[ D_{I {\rm boson}}(\vec p) D_{I {\rm boson}}(\vec k + \vec p) - D_{I {\rm fermion}}(\vec p) D_{I {\rm fermion}}(\vec k + \vec p)
 \right]_{\rm energy \ divergence} = 0.
 \eeq
 II.  Neglect loops with two non-instantaneous propagators,
 \beq
\int d^{d+1}p \ \left[ D_N(p) D_N(k + p)
 \right] \to 0.
 \eeq
III.  Keep loops with one instantaneous propagator and one non-instantaneous propagator,
 \beq
\int d^{d+1}p \ \left[ D_N(\vec p, p_0) D_I(\vec k + \vec p)
 \right].
 \eeq
Only the equal-time part of any non-instantaneous propagator contributes to the graphs we consider,
 \beq
\int d^{d+1}p \ \left[ D_N(\vec p, p_0) D_I(\vec k + \vec p) \right] = \int d^d p \ \left[ D^{\rm ET}(\vec p) D_I(\vec k + \vec p) \right],
 \eeq
where the equal-time part of the propagator is defined by
 \beq
\int dp_0 \ D_N(\vec p, p_0) = D^{\rm ET}(\vec p).
 \eeq
Thus, in all graphs that we consider, the non-instantaneous propagator gets replaced by its equal-time part.  The diagrams corresponding to the ISDE are given in Fig.~\ref{equations}.  
 
Calculations with the ISDE will be reported elsewhere \cite{Cooper:2016}.

\section{Conclusion}

We have a local, renormalizable quantum field theory with the following interesting properties:\\
$\bullet$  It provides a cut-off at the Gribov horizon.\\
$\bullet$ The Kugo-Ojima color confinement condition is satisfied.\\
$\bullet$ The vacuum is a perfect dielectric.\\
$\bullet$ The Maggiore-Schaden shift provides a BRST-invariant Lagangian.\\
$\bullet$ BRST-symmetry is spontaneously broken, but perhaps only in the unphysical sector.\\
$\bullet$ In Coulomb gauge, the color-Coulomb potential rises linearly or super-linearly when the Wilson potential is linearly rising.\\
$\bullet$  In the ISDE, all loops consist of one instantaneous propagator and the equal-time part of a non-instantaneous propagator.

%

\begin{thebibliography}{}
%
%
\bibitem{Gribov:1977wm}
V.N.~Gribov, 
Quantization of non-Abelian Gauge Theories,
Nucl. Phys. B {\bf 139}, 1 (1978).

\bibitem{Zwanziger:2002}
Daniel Zwanziger,
No Confinement without Coulomb Confinement,
Phys. Rev. Letts, {\bf 90,} 102001 (2003) and arXiv:hep-lat/0209105.

\bibitem{Zwanziger:1998}
Daniel Zwanziger, 
Renormalization in the Coulomb gauge and order parameter for confinement in QCD,
Nucl. Phys. B {\bf 518} 237-272 (1998).

\bibitem{Zwanziger:1989mf}
Daniel Zwanziger,
Local and Renormalizable Action from the Gribov Horizon,
Nucl.~Phys.~B{\bf 323}, 513 (1989).

\bibitem{Schaden:1994}
N.~Maggiore, M.~Schaden,
Landau gauge within the Gribov horizon,
Phys.~Rev.~D{\bf 50}, 6616 (1994).

\bibitem{Schaden:1996}
M.~Schaden, habilitation thesis on \emph{Kovariante Eichfixierung und Dynamische Brechung der BRS Symmetrie von Yang-Mills Eichtheorien}, Technische Universit{\"a}t M{\"u}nchen, Germany  (1996).

\bibitem{Schaden:1412}
Martin Schaden and Daniel Zwanziger,
Phys. Rev. {\bf D92,} 025001 (2015), and arXiv:1412.4823 [hep-ph].

\bibitem{Schaden:1501}
Martin Schaden and Daniel Zwanziger,
Living with spontaneously broken BRST symmetry. II. Poincar\'e invariance,
Phys.\ Rev.\ {\bf D92,} 025002 (2015) and arXiv:1501.05974 [hep-th].

\bibitem{Alkofer:2009dm}
Alkofer, Reinhard and Maas, Axel and Zwanziger, Daniel,
Truncating first-order Dyson-Schwinger equations in Coulomb-Gauge Yang-Mills theory,
Few Body Syst, {\bf 47}, 73 (2010) and arXiv:0905.4594.

\bibitem{Kugo-Ojima}
T.~Kugo, I.~Ojima, 
Prog.~Theor.~Phys.~Suppl.{\bf 66}, 1 (1979) ;Erratum Prog.~Theor.~Phys.{\bf 71}, 1121 (1984).

\bibitem{Kugo:1995km} 
T.~Kugo,
 The Universal renormalization factors Z(1) / Z(3) and color confinement condition in nonAbelian gauge theory, arXiv:hep-th/9511033.
 
 \bibitem{Reinhardt:2008}
H. Reinhardt, Phys. Rev. Lett. 101, 061602 (2008), 08030504.

\bibitem{Vandersickel:2012}
N.~Vandersickel, D.~Zwanziger, 
The Gribov problem and QCD dynamics,
Phys.~Rept.{\bf 520}, 175 (2012) and arXiv:1202.1491 [hep-th].

\bibitem{Sobreira:2004}
R.F.~Sobreira, S.P.~Sorella,
Introduction to the Gribov Ambiguities In Euclidean Yang-Mills Theories,
arXiv:hep-th/0504095.

\bibitem{Andrasi:2014}
A. Andrasi, J. C. Taylor,
The effective action in QCD,
Annals of Physics {\bf 351} 407-417 (2014) and arXiv:1406.7802 [hep-th], and references found there.

\bibitem{Cooper:1512.08}
Patrick Cooper and Daniel Zwanziger,
Local QCD Action at Finite Temperature,
Phys. Rev. {\bf D 93,} 105026 (2016) and arXiv:1512.08858.

\bibitem{Cooper:2016}
Patrick Cooper and Daniel Zwanziger (in preparation).



  
        





\end{thebibliography}
%
%


\end{document}